\documentclass[preprint,review,12pt,sort&compress]{elsarticle}

\usepackage{amssymb}
\usepackage{amsmath}

\journal{(not declared)}

\begin{document}

\begin{frontmatter}

\title{Metamaterial-induced-transparency engineering through quasi-bound states in the continuum by using dielectric cross-shaped trimers}

\author[label1]{Maryam Ghahremani}
\ead{ghahremani.maryam@ut.ac.ir}

\author[label2]{Carlos J. Zapata-Rodr\'{\i}guez\corref{corrau}}
\ead{carlos.zapata@uv.es}
\cortext[corrau]{Corresponding author}

\affiliation[label1]{
	organization={Photonics Research Laboratory, Center of Excellence on Applied Electromagnetic Systems, School of Electrical and Computer Engineering, College of Engineering, University of Tehran},
	city={Tehran},
    country={Iran}
    }

\affiliation[label2]{
	organization={Department of Optics and Optometry and Vision Sciences, Faculty of Physics, University of Valencia},
	addressline={Dr. Moliner 50}, 
	city={Burjassot},
	postcode={46100}, 
	country={Spain}
}

\begin{abstract}
	
This study presents a novel approach to activate a narrowband transparency line within a reflecting broadband window in all-dielectric metasurfaces, in analogy to the electromagnetically-induced transparency effect, by means of a quasi-bound state in the continuum (qBIC). 
We demonstrate that the resonance overlapping of a bright mode and a qBIC-based nearly-dark mode with distinct Q-factor can be fully governed by a silicon trimer-based unit cell with broken-inversion-symmetry cross shape, thus providing the required response under normal incidence of a linearly-polarized light. 
Our analysis that is derived from the far-field multipolar decomposition and near-field electromagnetic distributions uncovers the main contributions of different multipoles on the qBIC resonance, with governing magnetic dipole and electric quadrupole terms supplied by distinct parts of the dielectric ``molecule.'' 
The findings extracted from this research open up new avenues for the development of polarization-dependent technologies, with particular interest in its capabilities for sensing and biosensing.

\end{abstract}


\end{frontmatter}

\section{Introduction}

Metasurfaces are two-dimensional metamaterials with subwavelength thickness, which are renowned for their exceptional ability to manipulate waves.
This characteristic stems from their strong interaction with electric and/or magnetic fields~\cite{Sautter15,Genevet17,Hsiao17,Li18}.  
This flat-optics interaction is typically facilitated by resonant effects, which are mostly controlled by the geometry of the unit cells.
Thus, metasurfaces are capable of generating a variety of unique electromagnetic responses. 
These include anapole mode~\cite{Algorri18,Baryshnikova19,Ghahremani21}, Fano resonance~\cite{Cong15,Campione16,Limonov17}, electromagnetically-induced transparency~\cite{Yang14,Yahiaoui18,Diao19}, and bound states in the continuum (BICs)~\cite{Li19,Meng22,Mohamed22,Xiao22}, the latter are here devoting our attention. 
The unique capabilities of metasurfaces have led to planar optical meta-components with a wide range of applications across both microwave and optical frequencies in the fields of imaging and lensing~\cite{Hashemi16,Khorasaninejad17,Shanei17,Shrestha18}, holography~\cite{Ni13,Zheng15,Wen15}, structural color printing~\cite{Cheng15b,Sun17,Yang20}, energy harvesting~\cite{Zhang17,Ghaderi18,Li20}, and perfect absorbers~\cite{Akselrod15,Raad20,Karimi22}, among others. 
Importantly, the recent development of all-dielectric metasurfaces especially benefit from highly efficient transmission applications~\cite{Jahani16,Kamali18,Koshelev20}. 

Bound states in the continuum (BICs) are garnering increasing attention in the field of photonics due to their theoretically infinite quality factor (Q factor) and remarkable local field enhancement. 
In particular, symmetry-protected BICs are states that remain perfectly localized in dynamic systems even though they coexist with a continuous spectrum of radiation due to the symmetry mismatch between the bound state and free space.
The continually narrow linewidth of an ideal BIC makes it invisible in transmission spectra.
However BICs can degenerate into qBICs, typically in the form of a Fano transmission line, which exhibits a finite Q factor when the in-plane symmetry of the periodic metasurface is disrupted.
In this way, qBICs are allowed to couple to the radiative channel. 
The exceptional properties of a qBIC have found wide-ranging applications in various fields such as nonlinear optics~\cite{Liu19,Koshelev19,Zograf22,Kim24,Wang24}, lasers~\cite{Kodigala17,Spagele21,Tripathi21}, polarization conversion~\cite{Gorkunov20,Chen21,Pura22}, and sensors~\cite{Tan21,Wang21,Liu23,Hong24}, to name a few.

In recent years, extensive and comprehensive research has been conducted on electromagnetically-induced transparency (EIT) as a result of quantum interference; 
see for instance the review of Fleischhauer~\emph{et al.}~\cite{Fleischhauer05} and references therein.
In this context, the phenomenon of EIT is experienced in a medium that would typically be opaque but it generates a transparency range within a narrow bandwidth enabled by a light coupling resonance.
Lately, there has been a growing focus on the EIT effect concerning metamaterials, which is achieved through the coupling of bright and dark modes~\cite{Papasimakis08,Yang14}. 
Furthermore, in order to attain a narrowband transmission resonance, a pair of resonances is required exhibiting minimal interference offset and a substantial $Q$-factor contrast.

Despite its progress, research is still sparse on the transition of a symmetry-protected BIC into a perturbed-unit-cell qBIC for the excitation of a high-$Q$ EIT effect. 
Restricting our analysis to all-dielectric nanostructures, different proposals have been introduced enabling to control the spectral position of qBICs resonances to achieve a EIT response at normal incidence;
for a proposal based on oblique incidence see Ref.~\cite{Abujetas21}.
A first group of studies are focused on the coupling of a pair of dielectric resonators each one carrying either a bright mode or a dark mode~\cite{He21,Zheng24}.
In particular, the dark mode is activated by near-field coupling driven by the resonator which exhibits the bright mode.
A later proposal was based on the use of a single complex unit cell, such as for instance a square-slot ring~\cite{Algorri22}, or a tetrameric cluster~\cite{Huang22}.
While previous research has primarily focused on a single EIT and its applications in slow light effect or refractive index sensing, it has overlooked the potential for simultaneous excitation of high-Q EIT effect and qBIC modes on proposed metasurfaces, as well as the polarization-dependent response provided the in-plane symmetry perturbation required for their excitation always breaks the inversion symmetry of the structures.

In this paper, polarization-sensitive qBIC resonance is shown in a dielectric cross-shaped trimer metasurface. 
This complex metasurface exhibits a bright mode controlled by a single Si bar, which is set along the direction of polarization of the impinging light, and a nearly dark mode governed by broken inversion-symmetry dimer yet-connected with perpendicular orientation upon polarization.
The trimer being physically connected makes the difference from other dielectric and plasmonic proposals~\cite{Zhang08c,Liu09}.
Furthermore, the qBIC resonance emerges when the in-plane symmetry is violated. 
Driven by the main electric quadrupole and magnetic dipole dependence of the qBIC, the $Q$ factor and resonant wavelength of its transmittance peak leading to an EIT-like effect is controlled by relative displacements of the dimer.

\section{Structure design}

\begin{figure}[t] 
	\includegraphics[width=0.60\textwidth]{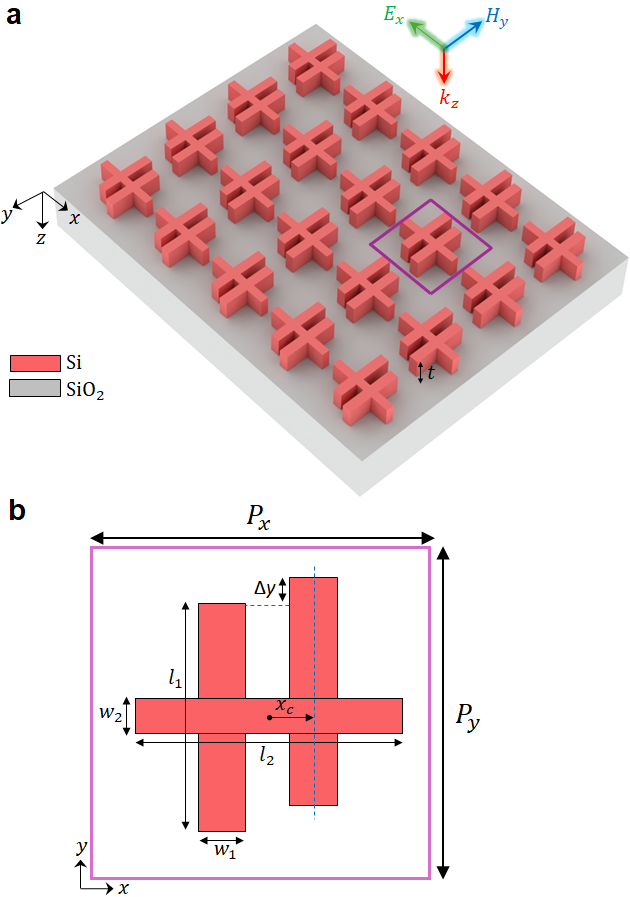}
	\centering
	\caption{
		Schematics of (a) the metasurface, (b) a single unit cell including the dielectric ``meta-molecule.''
	}
	\label{fig:01}
\end{figure}

In Fig.~\ref{fig:01}, we show a metasurface composed of a square lattice of silicon cross-shaped bar trimers deposited on a SiO$_2$ substrate lying on the $xy$ plane. 
The ``meta-molecule'' is composed of a pair of identical (vertical) bars oriented along the $y$ axis and a long horizontal Si bar cutting across the above-mentioned dimer symmetrically.
Note the overlapping volume existing between both substructures, which latter will demonstrate a key role in the optical response of the nanostructure.
A plane wave impinges upon the metasurface along the negative $z$-axis direction.
Structural parameters for a single unit cell are detailed in the inset of Fig.~\ref{fig:01}. 
The substrate height $h_\text{Si}$ measures 190~nm, while the period $P_x = P_y$ of the squared lattice reaches 780~nm. 
The bar dimer is specifically designed with a length $l_1$ of 550~nm and a side length $w_1$ for each individual bar at 95~nm. 
In this symmetrical configuration, the gap $2 x_c$ between adjacent bars as measured from their centers is 170~nm.
Note that by changing the variable $x_c$ we assume that vertical bars shift symmetrically from the center of the horizontal bar.
The horizontal Si bar has a length $l_2 = 650$~nm and a width $w_2 = 80$~nm.
The height $t$ of the Si bars is $190$~nm.
We achieve qBIC resonances by simply displacing the right vertical bar a length $\Delta y$ along the $y$-axis.
The state of polarization of the incident wave plays a role, since plane-wave polarization along the $y$-axis reveals an inert action in the formation of a qBIC.
We quantify the system asymmetry using the parameter
\begin{equation} \label{eq:02}
	\alpha
	=
	\frac{\Delta y}{l_1} ,
\end{equation}
which stands no more than for a normalized displacement of the right vertical Si bar.
For the sake of symmetry, we will consider positive values of $\Delta y$, providing positive asymmetry parameters. 
The refractive indices for silicon and the substrate are set at 3.5 and 1.45, respectively.

\section{Results and discussion}

\begin{figure}[t] 
	\includegraphics[width=0.60\textwidth]{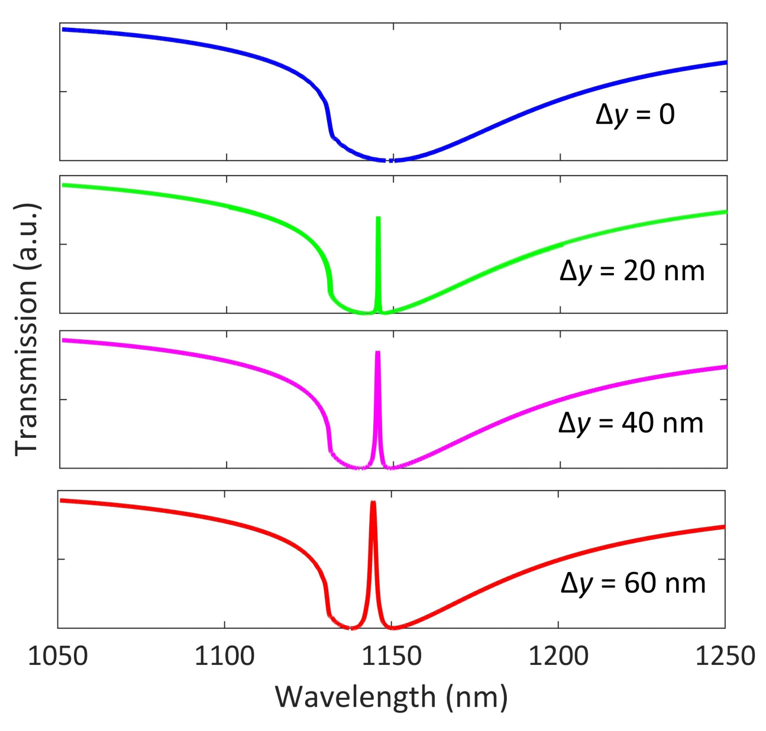}
	\centering
	\caption{
		Transmission spectra of the metasurface with increasing $\Delta y$.  
		The geometrical parameters of the Si bars are: $w_1 = 95$~nm, $w_2 = 80$~nm, $t = 190$~nm, $l_1 = 550$~nm, $l_2 = 650$~nm.
	}
	\label{fig:02}
\end{figure}

To analyze the metasurface behavior when varying different geometrical parameters of the unit cell, we simulate its transmission spectra applying the finite-difference time-domain (FDTD) method through the Lumerical FDTD package, as frequencly encountered elsewhere~\cite{Karimi22}.
Some of the resultant transmittances within the spectral range of interest are depicted in Fig.~\ref{fig:02}, where we vary the parameter $\Delta y$ from 0 to 60~nm under $x$-linearly polarized illumination at normal incidence.
At $\Delta y = 0$~nm, the transmission spectrum spanning from 1050~nm to 1250~nm shows a broadband resonant minimum with nearly 100\% of reflectance at a wavelength of 1146~nm. 
From 1140~nm to 1160~nm transmission remains nearly flat, exhibiting a plateau of nearly-zero transmittance. 
As we will see below, a bound state in the continuum (BIC) can be found within this spectral band, however, remaining invisible in the transmission spectrum.
In this regards, BICs manifest a resonance with an infinitely narrow linewidth, resulting in an infinitely high $Q$ factor. 
Nevertheless, as we increase $\Delta y$, the symmetry of the structure is disrupted, leading to the emergence of a BIC-related resonance in the transmission spectrum. 
As shown in Fig.~\ref{fig:02}, the peak transmittance becomes broader as long as $\Delta y$ (and thus the asymmetry parameter $\alpha$) is increasing, however the resonant frequency remains practically invariant.
This behavior leading to the analogous of the EIT effect can be attributed to the interaction between the bound state and the radiative channel, subsequently exciting a qBIC resonance. 

\begin{figure*}[t] 
	\includegraphics[width=0.95\textwidth]{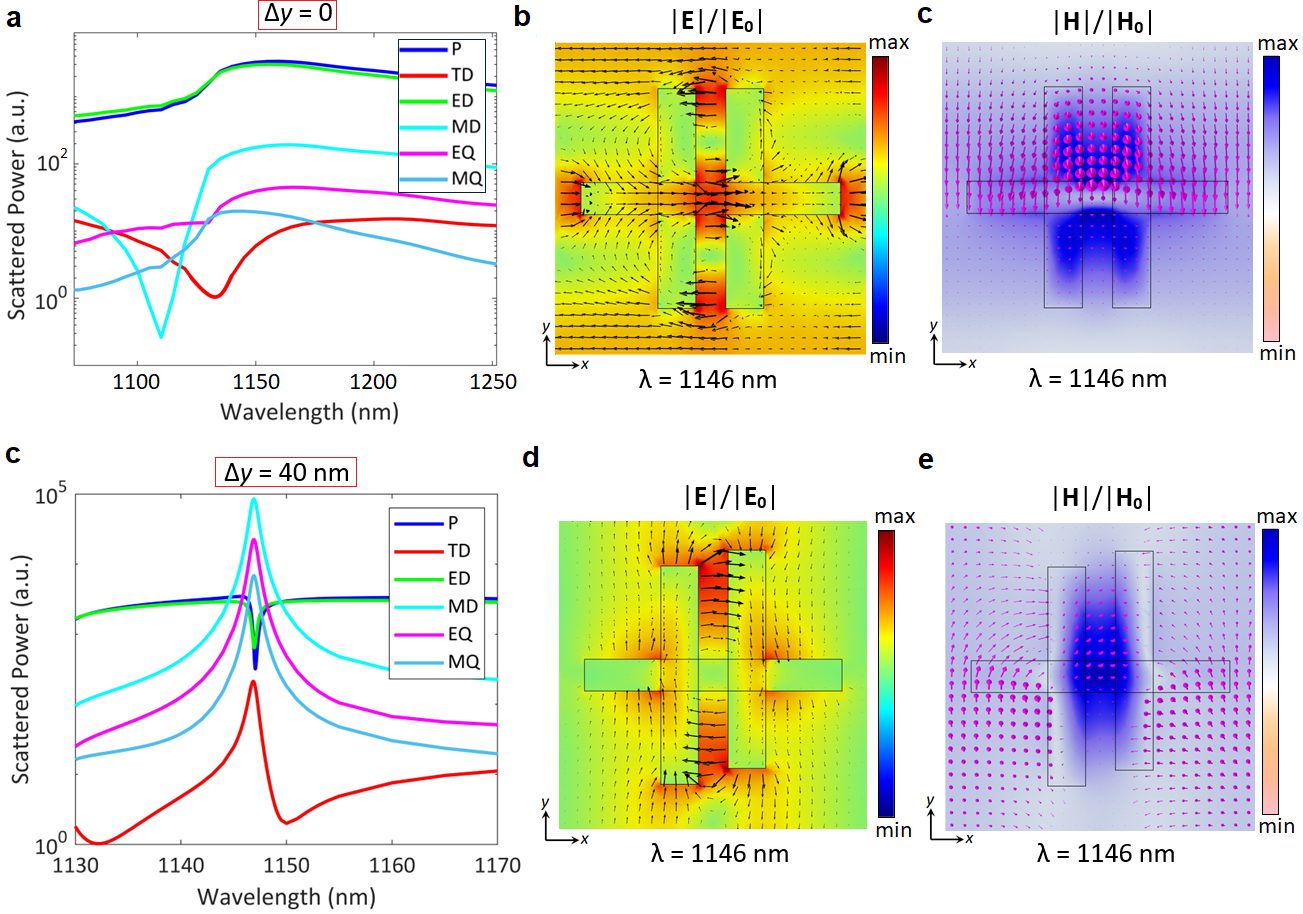}
	\centering
	\caption{
		Electric (a) and magnetic (b) field distribution of the resonance when $\Delta y = 0$~nm at a wavelength of $\lambda = 1132$~nm.
		(c) Multipolar decompositions for the qBIC for the symmetric nanostructure.
		Figures~(d)-(f) refer to a displacement $\Delta y = 40$~nm.
	}
	\label{fig:03}
\end{figure*}

Next the analysis of near-field electromagnetic field distributions is conducted to elucidate the mechanism underlying the EIT observed about the qBIC. 
For comparative purposes, Fig.~\ref{fig:03} (a) and (b) illustrates the electric field and magnetic field pattern in the $xy$ plane at a wavelength of $\lambda = 1146$~nm for the symmetric structure set at $\Delta y = 0$. 
For completeness, the multipoles scattered power in the far field is determined through multipolar decomposition, which is graphically represented in Fig.~\ref{fig:03}(c).
The mathematical treatment here followed for the evaluation of the scatterered power under the multipolar analysis is given in~\ref{sec:appendix}.
Notably, the contribution of the electric dipole (ED) predominates in the full optical response, overshadowing the minor contributions from the magnetic dipole (MD) and electric quadrupole (EQ). 
In the mirror-symmetric arrangement, the electric field is mainly localized along the horizontal Si bar and pointing along the $x$ axis, thus being parallel to the incident polarized field.
This electric dipolar interaction of the field with the nanostructure with net radiative emission along the $z$ axis, thus representing a bright mode, is responsible for the broadband minimum in the metasurface transmission discussed above.
On the other hand, when the diplacement $\Delta y$ increases to 40~nm, the electric field is mainly distributed around the top and botton bases of the bar dimmer and the space in between them, while the magnetic field predominantly resides within the structure center.
They exhibit characteristic features of an EQ and MD instead, as indicated by the black arrows in Fig.~\ref{fig:03}(d) and (e).
The latter is confirmed by the multipolar analysis displayed in Fig.~\ref{fig:03}(f).
Importantly, both EQ and MD resonances are radiationless along the $z$ direction, which is characteristic of a dark mode, enabling the appearance of a narrowband trnasmission line at the centered wavelength of $1146$~nm.

\begin{figure}[t] 
	\includegraphics[width=0.45\textwidth]{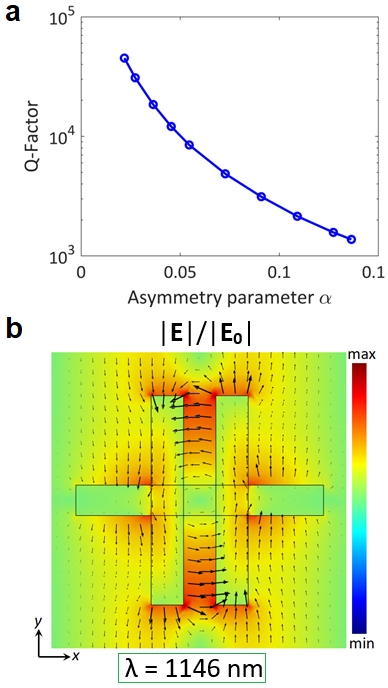}
	\centering
	\caption{
		(a) Variations in the $Q$ factors for qBIC with increasing asymmetry. 
		(b) Resonant field of a BIC with a natural wavelength of $1146$~nm found for the symmetric configuration.
	}
	\label{fig:04}
\end{figure}

\begin{figure*}[t] 
	\includegraphics[width=0.95\textwidth]{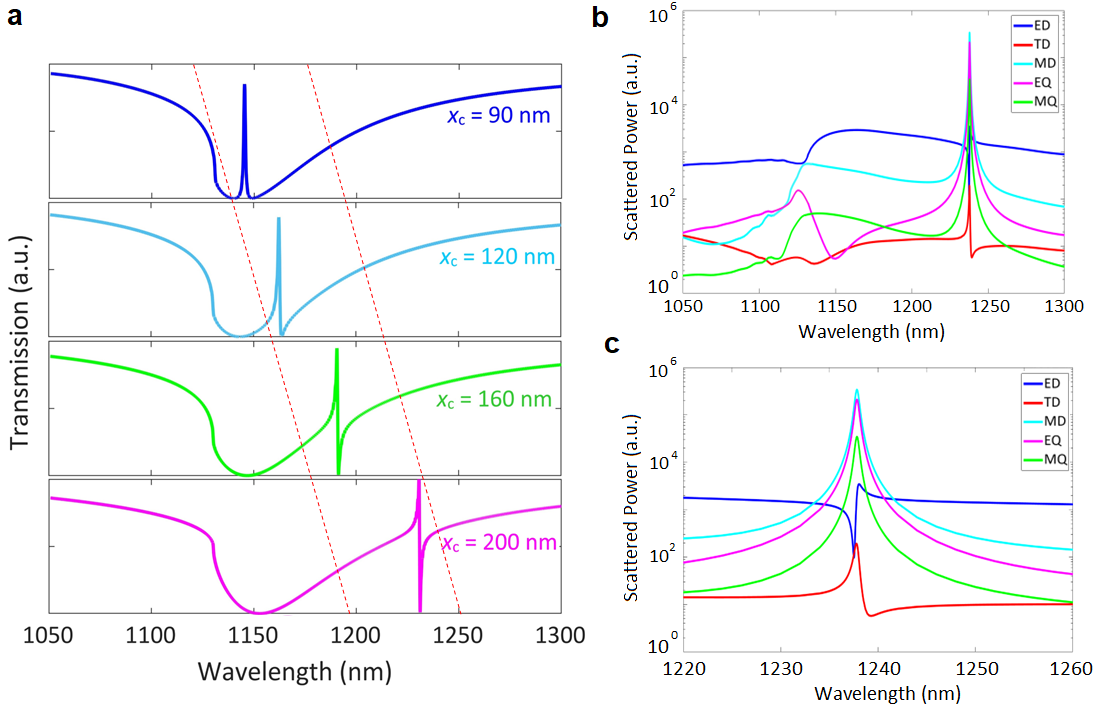}
	\centering
	\caption{
		(a) Transmission spectra of the metasurface with increasing $\Delta y = 40$~nm and different values of $x_c$. 
		(b) and (c) show the multipolar decomposition for the qBIC for the asymmetric nanostructure at $x_c = 200$~nm.
	}
	\label{fig:05}
\end{figure*}

To further analyze the qBIC nature of the narrowband resonance around $\lambda = 1146$~nm we evaluate the complex-valued eigenfrequencies of our nanostructure when varying the displacement $\Delta y$ and thus the asymmetry parameter $\alpha$.
For that purpose, we use the finite element method (COMSOL Multiphysics) by applying periodic boundary conditions along the $x$ and $y$ directions, thus encircling the metasurface unit cell, however including a perfectly-matched layer at the boundaries of the computing domain along the $z$ direction~\cite{Han19,Han21,Chai21,You23,Ghahremani23}.
The $Q$-factor for our qBIC resonance can be mathematically expressed as 
\begin{equation}
	Q 
	= 
	\frac{\operatorname{Re} \omega}{2 \operatorname{Im} \omega} ,
\end{equation}
provided the imaginary part of the eigenfrequency does not vanishes.
In this concern, a pure BIC is revealed by the condition $\operatorname{Im} \omega = 0$, or equivalent by an infinitely high $Q$ factor, thus denoting a nonradiative wave within the metasurface.
Note that the numerical evaluation of the $Q$ factors can alternatively benefits from the Fano model, which allows to fit the spectral features of the transmittance pattern~\cite{Limonov17}; 
importantly, both methods are in full agreement~\cite{Han19}.
Examining the behavior of the qBIC resonance, we observe changes in its $Q$ factor as the system asymmetry increases. 
Specifically, as the asymmetry gradually grows, the $Q$ factor of this resonance decreases. 
The relationship between the $Q$ factor of a qBIC resonance and the asymmetry parameter $\alpha$ can be quantified as $Q \propto \alpha^{-2}$. 
Furthermore, in Fig.~\ref{fig:04}(a), we visualize the variations in the $Q$ factors with increasing asymmetry. 
For $\Delta = 0$ we find a quality factor $Q \to \infty$, revealing the existence of a BIC.
Quality factors exceeding $10^4$ can be achieved for values of the asymmetry parameter $\alpha$ lower that $1/20$.
The electric field of such a bound mode is shown in Fig.~\ref{fig:04}(b), which present similar features as the resonant field shown in Fig.~\ref{fig:03}(d) for the qBIC at $\Delta y = 40$~nm.
Overall, the findings in Figs.~\ref{fig:04} demonstrates that the qBIC is a symmetrically protected qBIC resonance. 

While the vertical displacement $\Delta y$ governs the $Q$ factor of the symmetrically-protected qBIC, as explained above, the horizontal displacement $x_c$ of the right-sided bar determines the resonant wavelength of the transparency peak.
Figure~\ref{fig:05}(a) shows the metasurface transmission for a fixed vertical displacement $\Delta y = 40$~nm and different horizontal displacement $x_c$.
The transmission peak is redshifted from $1150$~nm at a displacement $x_c$ of 90~nm to a resonance at $1230$~nm if $x_c = 200$~nm.
Within such a range of horizontal displacements, the background broadband ED resonance remains practically unaltered due to the fact that it is controlled by the horizontal Si bar.
Again, an interaction can be observed between the bright ED broadband resonance and a dark narrowband resonance contolled mainly by a MD and EQ moment contributions in the multipolar analysis, as shown in Fig.~\ref{fig:05}(b) and (c).
More especifically, the Fano model well fits the spectral features as~\cite{Limonov17,Cai21,Fan22},
\begin{equation} \label{eq:01}
	T (\omega)
	=
	T_0
	+
	A_0
	\frac{[q + 2 (\omega - \omega_0)/\gamma]^2}{1 + [2 (\omega - \omega_0)/\gamma]^2} ,
\end{equation}
where $\omega_0$ is the resonant frequency, $\gamma$ is the resonance linewidth, and $T_0$ is the transmission offset, $A_0$ is the continuum-discrete coupling constant, $q$ is the Breit-Wigner-Fano parameter determining asymmetry of the resonance profile. 
Anyhow, for high values of $x_c$, the transparency peak occurs at a non-negligible background signal, leading to a Fano shape in the transmission pattern.

\begin{figure*}[t] 
	\includegraphics[width=0.75\textwidth]{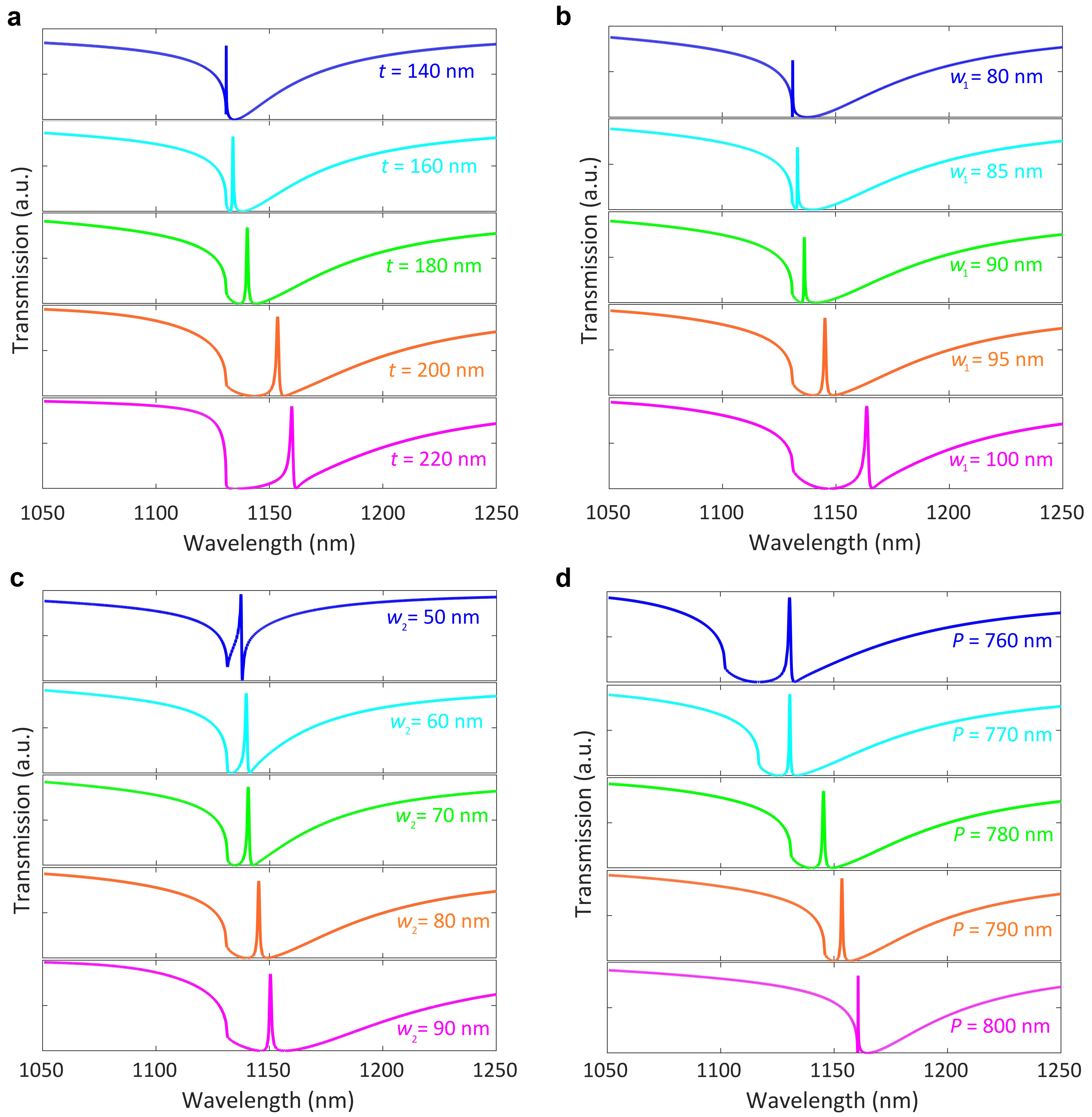}
	\centering
	\caption{
		Transmission spectra of the metasurface with varying different geometrical parameters: the height of the cross-shaped resonators $t$ (a), the bar width $w_1$ (b) and $w_2$ (c), and the lattice period $P$ (d).
		At each panel, the remaining parameters are fixed to the values assigned to the design described in Fig.~\ref{fig:02}. 
	}
	\label{fig:06}
\end{figure*}

Figure~\ref{fig:06} shows the metasurface transmission for varying geometrical parameters such as the height of the Si trimer, $t$, the width of the Si bars $w_1$ and $w_2$ and the lattice period $P$.
Notably, the latter is the only parameter enabling to shift the bright-mode ED resonance.
In particular, a redshift of the transmittance minimum is experienced for growing values of $P$, provided the unit-cell geometrical parameters remain fixed;
a minimum around the wavelength of 1120~nm occurs for a period of 760~nm, whereas the resonant wavelength rises to 1170~nm when $P = 800$~nm.
As a consequence, this becomes the optical mechanism to control the spectral window where EIT will apply.
On the other hand, increasing values of either $t$, $w_1$ or $w_2$ keep practically unaltered the spectral response of the bright mode, however achieving a redshift of the EIT peak in all cases.
One might expect that the redshift were certainly short for width variations of the horizontal Si bar, which mainly controls the bright mode, as numerically confirmed in the simulations shown in Fig.~\ref{fig:06}(c).
As a results, higher redshift rates can be achieved when changes of the parameters $t$ and $w_1$ are executed.
The latter case involving the bar width of the Si dimer oriented along the $y$ axis, which strongly governs the dark mode behind the BIC, demonstrate how sensitive this procedure is, enabling a feasible control of the peak frequency where transparency might be required.

\section{Application: refractive-index sensing}

High $Q$-factor Fano resonances can be excited using all-dielectric nanostructures made of low-loss, high refractive index materials.  
It will offer unexpected concepts for implementing highly integrated, miniaturized, and high-performance photonic devices, which have promising uses in environmental monitoring and sensing~\cite{Wang21b,Hsiao22,Algorri23,Zhou23}.
Similarly, the analogous of EIT in metasurfaces can serve as the basis for the design of all-dielectric optical sensors.
However, such application has only been numerically explored when the EIT is enabled by qBICs~\cite{Algorri22,Huang22,Li23,Wang23}.

\begin{figure*}[t] 
	\includegraphics[width=0.95\textwidth]{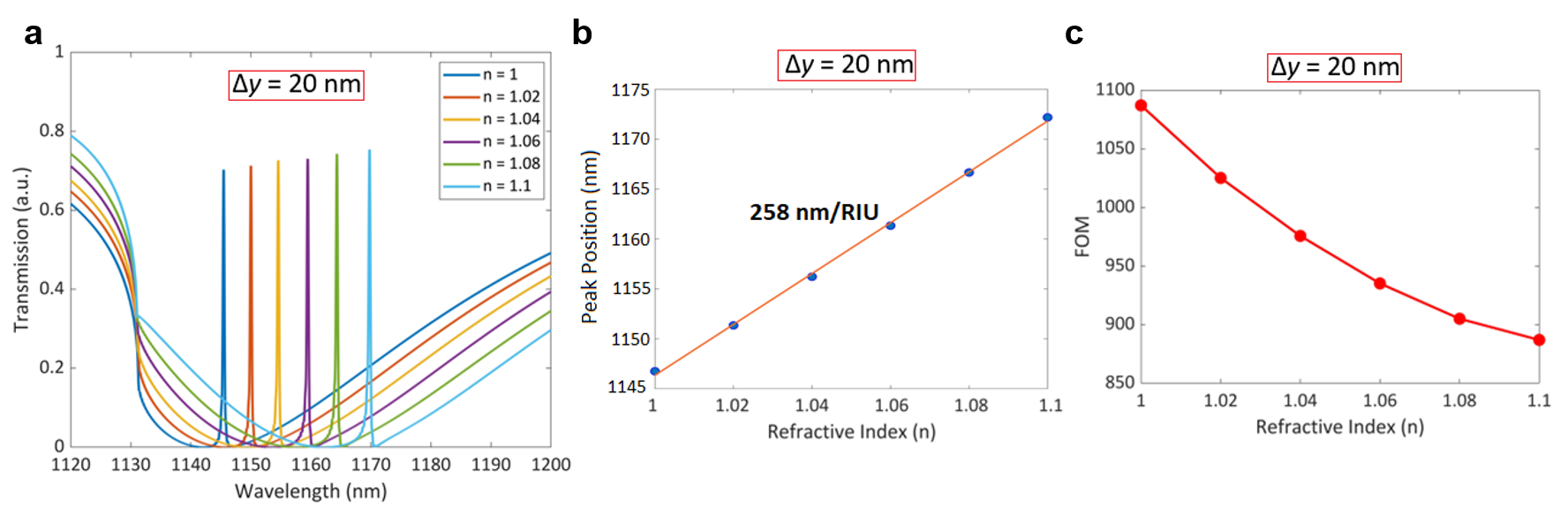}
	\centering
	\caption{
		(a) Transmission spectra of the metasurface with varying the index of refraction of the environment medium.
		(b) Transparency peak wavelength and (c) FOM vs the index of refraction of the environment medium as estimated for a metasurface with $\Delta y = 20$~nm.
	}
	\label{fig:07}
\end{figure*}

We have examined how the transmission spectrum varies in a different ambient refractive index $n$ based on the features of our design provided above. 
Here we set $\Delta y = 20$~nm.
First, we focus on the peak wavelength of the induced transparency when $n$ varies in steps of 0.02, which is depicted in Fig.~\ref{fig:07}(a) and (b), from which it is evident that the resonant wavelength redshift with increasing refractive index. 
This is a normal behavior that has been observed when using qBICs elsewhere~\cite{Algorri23,Huang22,Li23}.
The sensitivity $S$ can be used to measure the sensor performance. 
The sensitivity $S$ is defined as the variation of the resonant wavelength per unit refractive index (RIU).
Thus, the following formula can be used to estimate the sensitivity: $S = \Delta \lambda / \Delta n$, typically given in units of nm/RIU, where $\Delta n$ is the difference in refractive index and $\Delta \lambda$ is the corresponding wavelength offset. 
In our simulations, this value reaches 258~nm/RIU within the range $1 \le n \le 1.1$ for the architecture characterized by s spatial shift $\Delta y$ of 20~nm.
This value is higher than the sensitivity, or in the same order, as exhibited by previous proposals based on EIT driven by qBICs~\cite{Algorri22,Huang22,Li23}.
In addition, a figure of merit (FOM) is commonly included in the valoration of the sensor performance.
The formula for calculating the FOM value is~\cite{Sherry05,Yang14,Huang22}:
\begin{equation}
	\text{FOM}
	=
	\frac{S (\text{nm/RIU})}{\Delta (\text{nm})} ,
\end{equation}
where $\Delta$ is the full width between the transmission peak wavelength and the transmission dip wavelength.
As a result, as illustrated in Fig.~\ref{fig:07}(b), we compute the resonance wavelength under various refractive indices. 
FOM values can surpass $1 \times 10^3$ RIU$^{-1}$, which represents a notably large value, even higher by one order of magnitude in comparison with previous propsals~\cite{Algorri22,Huang22,Li23}.

\begin{figure}[t] 
	\includegraphics[width=0.45\textwidth]{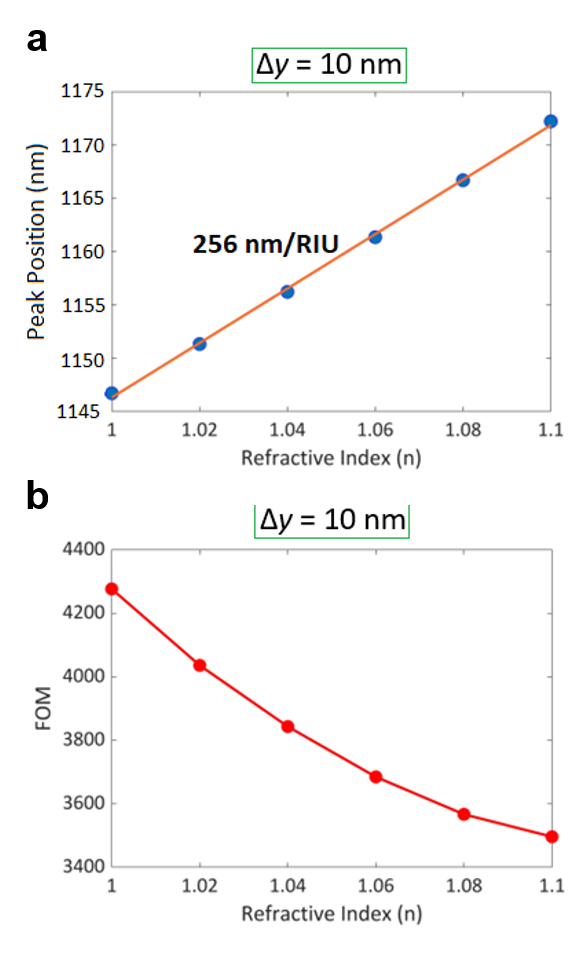}
	\centering
	\caption{
		The same as Fig.~\ref{fig:07} but using $\Delta y = 10$~nm. 
	}
	\label{fig:08}
\end{figure}

For the sake of completeness we performed an analysis of the sensitivity and FOM for an asymmetric metasurface with $\Delta y = 10$~nm.
Both photonic metasurfaces with an asymmetric parameter of 20~nm and 10~nm, respectively, have in practice the same sensitivity.
More specifically $S$ gets slightly reduced to a sensitivity of 256~nm/RIU, as seen in Fig.~\ref{fig:08}(a).
However, $\Delta y = 10$~nm provides higher $Q$ factors and thus a higher FOM, as shown in Fig.~\ref{fig:08}(b).
Note that $\text{FOM} = S Q / \lambda_\textbf{qBIC}$, where $\lambda_\textbf{qBIC}$ and 
$Q$ denote the resonant wavelength and $Q$ factor of the qBIC~\cite{Li23}, justifying that FOM greatly increases by a factor of 4.
This is an expected behavior since the $Q$ factor of the resonances changes upon the asymmetry factor $\alpha$ given in Eq.~\eqref{eq:02} by an inverse squared factor leading to the relation $Q \propto (\Delta y)^{-2}$, as discussed above.
On the contrary, an asymmetry of $\Delta y = 10$~nm leads to a lower transmittance, which reaches to a peak value of 0.6, in comparison with a peak transmittance of 0.8 for the case of $\Delta y = 20$~nm.
As a results, we find a trade-off between high values of the quality factor and the induced transmittance of the resonant qBIC.

\section{Conclusions}

In summary, with the coupling of a bright mode governed by a single dielectric bar and a dark mode controlled by a $90^\circ$-rotated asymmetric bar dimer, overall forming a cross-shaped Si resonator, an innovative approach for the excitation of a polarization-sensitive sharp resonance within a full-reflective background is presented. 
This scheme enables the formation of an analogue of the EIT effect.
It was revealed through the analysis of near-field electromagnetic distributions and far-field scattering the qBIC nature of this narrowband transmittance peak, whose origin is the in-plane symmetry break of the Si barred dimer with dominant MD and EQ multipolar contributions.
For varying relative displacements of the dimer components, the resonant mode can be reliably controlled, modifying the resonant wavelength and its $Q$ factor, albeit the respective contributions of the EQ and the MD to the polarization-sensitive qBIC remains unaltered. 
Our achievement enables it possible to overcome the restriction imposed by complex broken-symmetry geometries, affording a different approach for creating polarization-sensitive devices with qBIC resonances as well as promoting possible uses in a range of nanophotonic applications.
For the sake of clarity, we demonstrate its capabilities on the field of refractive-index sensing.

\section*{Funding}

This research received no external funding.

\appendix

\section{Multipolar analysis of qBICs}
\label{sec:appendix}

To elucidate the properties of the excited qBIC, we conducted near-field calculations through a multipolar analysis for the metasurface. 
In particular, we computed the scattering cross-section spectra of the Cartesian multipoles in free space by means of the following expressions~\cite{Tian16,Terekhov17,Zenin17,Babicheva21}:
\begin{align}
	\sigma_\mathrm{sca}
	\approx&
	\frac{k_0^4}{6 \pi \epsilon_0^2 |\mathbf{E}_\mathrm{inc}|^2}
	\left|
	\mathbf{P} + \frac{i k_0}{c} \mathbf{T}
	\right|^2
	+
	\frac{k_0^4 \mu_0}{6 \pi \epsilon_0 |\mathbf{E}_\mathrm{inc}|^2}
	\left|
	\mathbf{M} 
	\right|^2
	\nonumber
	\\
	&+
	\frac{k_0^6}{720 \pi \epsilon_0^2 |\mathbf{E}_\mathrm{inc}|^2}
	\sum_{\alpha \beta}
	\left|
	Q^{(e)}_{\alpha \beta}
	\right|^2
	\nonumber
	\\
	&+
	\frac{k_0^6 \mu_0}{80 \pi \epsilon_0 |\mathbf{E}_\mathrm{inc}|^2}
	\sum_{\alpha \beta}
	\left|
	Q^{(m)}_{\alpha \beta}
	\right|^2 ,
\end{align}
where $\epsilon_0$, $\mu_0$ and $c$ refers to electric permittivity, magnetic permeability and speed of light in free space, respectively, $\alpha, \beta = x, y, z$, $k_0$ is the vacuum wavenumber, and $\mathbf{E}_\mathrm{inc}$ stands for the incident electric field. 
The terms 
\begin{align}
	\mathbf{P}
	&=
	\frac{i}{\omega}
	\int_\Omega \mathbf{j} \text{d}v ,
	\\
	\mathbf{M}
	&=
	\frac{1}{2}
	\int_\Omega \mathbf{r} \times \mathbf{j} \text{d}v ,
	\\
	\mathbf{T}
	&=
	- \frac{1}{10}
	\int_\Omega 
	\left[
	2 r^2 \mathbf{j}
	-
	(\mathbf{r} \cdot \mathbf{j}) \mathbf{r}
	\right]
	\text{d}v ,
	\\
	Q_{\alpha \beta}^{(e)}
	&=
	\frac{3 i}{\omega} \int_\Omega 
	\left[ 
	{r}_\alpha {j}_\beta + {j}_\alpha {r}_\beta 
	-
	\frac{2}{3} (\mathbf{r} \cdot \mathbf{j}) \delta_{\alpha \beta} 
	\right]
	\text{d}v ,
	\\
	Q_{\alpha \beta}^{(m)}
	&=
	\frac{1}{3} \int_\Omega 
	\left[ 
	(\mathbf{r} \times \mathbf{j})_\alpha {r}_\beta 
	+ 
	{r}_\alpha (\mathbf{r} \times \mathbf{j})_\beta ,
	\right]
	\text{d}v ,
\end{align}
are the moments of electric dipole (ED), magnetic dipole (MD), toroidal dipole (TD), electric quadrupole (EQ), and magnetic quadrupole (MQ), respectively, which are evaluated in the spatial domain $\Omega$ of the scatterer through the induced displacement-related current density,
\begin{equation}
	\mathbf{j} (\mathbf{r})
	= 
	- i \omega \epsilon_0 
	\left[
	\epsilon (\mathbf{r}) - 1
	\right] 
	\mathbf{E} (\mathbf{r}) .
\end{equation}

\bibliographystyle{elsarticle-num} 

\begin{thebibliography}{10}
	\expandafter\ifx\csname url\endcsname\relax
	\def\url#1{\texttt{#1}}\fi
	\expandafter\ifx\csname urlprefix\endcsname\relax\def\urlprefix{URL }\fi
	\expandafter\ifx\csname href\endcsname\relax
	\def\href#1#2{#2} \def\path#1{#1}\fi
	
	\bibitem{Sautter15}
	J.~Sautter, I.~Staude, M.~Decker, E.~Rusak, D.~N. Neshev, I.~Brener, Y.~S.
	Kivshar, Active tuning of all-dielectric metasurfaces, ACS Nano 9~(4) (2015)
	4308--4315.
	
	\bibitem{Genevet17}
	P.~Genevet, F.~Capasso, F.~Aieta, M.~Khorasaninejad, R.~Devlin, Recent advances
	in planar optics: from plasmonic to dielectric metasurfaces, Optica 4~(1)
	(2017) 139--152.
	
	\bibitem{Hsiao17}
	H.-H. Hsiao, C.~H. Chu, D.~P. Tsai, Fundamentals and applications of
	metasurfaces, Small Methods 1~(4) (2017) 1600064.
	
	\bibitem{Li18}
	A.~Li, S.~Singh, D.~Sievenpiper, Metasurfaces and their applications,
	Nanophotonics 7~(6) (2018) 989--1011.
	
	\bibitem{Algorri18}
	J.~F. Algorri, D.~C. Zografopoulos, A.~Ferraro, B.~Garc{\'\i}a-C{\'a}mara,
	R.~Vergaz, R.~Beccherelli, J.~M. S{\'a}nchez-Pena, Anapole modes in hollow
	nanocuboid dielectric metasurfaces for refractometric sensing, Nanomaterials
	9~(1) (2018) 30.
	
	\bibitem{Baryshnikova19}
	K.~V. Baryshnikova, D.~A. Smirnova, B.~S. Luk'yanchuk, Y.~S. Kivshar, Optical
	anapoles: concepts and applications, Adv. Opt. Mater. 7~(14) (2019) 1801350.
	
	\bibitem{Ghahremani21}
	M.~Ghahremani, M.~K. Habil, C.~J. Zapata-Rodriguez, Anapole-assisted giant
	electric field enhancement for surface-enhanced coherent {anti-Stokes}
	{Raman} spectroscopy, Sci. Rep. 11~(1) (2021) 10639.
	
	\bibitem{Cong15}
	L.~Cong, M.~Manjappa, N.~Xu, I.~Al-Naib, W.~Zhang, R.~Singh, Fano resonances in
	terahertz metasurfaces: a figure of merit optimization, Adv. Opt. Mater.
	3~(11) (2015) 1537--1543.
	
	\bibitem{Campione16}
	S.~Campione, S.~Liu, L.~I. Basilio, L.~K. Warne, W.~L. Langston, T.~S. Luk,
	J.~R. Wendt, J.~L. Reno, G.~A. Keeler, I.~Brener, et~al., Broken symmetry
	dielectric resonators for high quality factor {Fano} metasurfaces, ACS
	Photonics 3~(12) (2016) 2362--2367.
	
	\bibitem{Limonov17}
	M.~F. Limonov, M.~V. Rybin, A.~N. Poddubny, Y.~S. Kivshar, Fano resonances in
	photonics, Nat. Photonics 11~(9) (2017) 543--554.
	
	\bibitem{Yang14}
	Y.~Yang, I.~I. Kravchenko, D.~P. Briggs, J.~Valentine, All-dielectric
	metasurface analogue of electromagnetically induced transparency, Nat.
	Commun. 5~(1) (2014) 5753.
	
	\bibitem{Yahiaoui18}
	R.~Yahiaoui, J.~A. Burrow, S.~M. Mekonen, A.~Sarangan, J.~Mathews, I.~Agha,
	T.~A. Searles, Electromagnetically induced transparency control in terahertz
	metasurfaces based on bright-bright mode coupling, Phys. Rev. B 97~(15)
	(2018) 155403.
	
	\bibitem{Diao19}
	J.~Diao, B.~Han, J.~Yin, X.~Li, T.~Lang, Z.~Hong, Analogue of
	electromagnetically induced transparency in an {S-shaped} all-dielectric
	metasurface, IEEE Photonics J. 11~(3) (2019) 1--10.
	
	\bibitem{Li19}
	S.~Li, C.~Zhou, T.~Liu, S.~Xiao, Symmetry-protected bound states in the
	continuum supported by all-dielectric metasurfaces, Phys. Rev. A 100~(6)
	(2019) 063803.
	
	\bibitem{Meng22}
	B.~Meng, J.~Wang, C.~Zhou, L.~Huang, Bound states in the continuum supported by
	silicon oligomer metasurfaces, Opt. Lett. 47~(6) (2022) 1549--1552.
	
	\bibitem{Mohamed22}
	S.~Mohamed, J.~Wang, H.~Rekola, J.~Heikkinen, B.~Asamoah, L.~Shi, T.~K. Hakala,
	Controlling topology and polarization state of lasing photonic bound states
	in continuum, Laser Photonics Rev. 16~(7) (2022) 2100574.
	
	\bibitem{Xiao22}
	X.~Xiao, Y.~Lu, J.~Jiang, Y.~Chen, Manipulation of optical bound states in the
	continuum in a metal-dielectric hybrid nanostructure, Photonics Res. 10~(11)
	(2022) 2526--2531.
	
	\bibitem{Hashemi16}
	M.~Hashemi, A.~Moazami, M.~Naserpour, C.~J. Zapata-Rodr{\'\i}guez, A broadband
	multifocal metalens in the terahertz frequency range, Opt. Commun. 370 (2016)
	306--310.
	
	\bibitem{Khorasaninejad17}
	M.~Khorasaninejad, F.~Capasso, Metalenses: versatile multifunctional photonic
	components, Science 358~(6367) (2017) eaam8100.
	
	\bibitem{Shanei17}
	M.~M. Shanei, M.~Hashemi, D.~Fathi, C.~J. Zapata-Rodr{\'\i}guez, Dielectric
	metalenses with engineered point spread function, Appl. Opt. 56~(32) (2017)
	8917--8923.
	
	\bibitem{Shrestha18}
	S.~Shrestha, A.~C. Overvig, M.~Lu, A.~Stein, N.~Yu, Broadband achromatic
	dielectric metalenses, Light Sci. Appl. 7~(1) (2018) 85.
	
	\bibitem{Ni13}
	X.~Ni, S.~Ishii, A.~V. Kildishev, V.~M. Shalaev, Ultra-thin, planar,
	{Babinet-inverted} plasmonic metalenses, Light Sci. Appl. 2~(4) (2013) e72.
	
	\bibitem{Zheng15}
	G.~Zheng, H.~M{\"u}hlenbernd, M.~Kenney, G.~Li, T.~Zentgraf, S.~Zhang,
	Metasurface holograms reaching 80\% efficiency, Nat. Nanotechnol. 10~(4)
	(2015) 308--312.
	
	\bibitem{Wen15}
	D.~Wen, F.~Yue, G.~Li, G.~Zheng, K.~Chan, S.~Chen, M.~Chen, K.~F. Li, P.~W.~H.
	Wong, K.~W. Cheah, et~al., Helicity multiplexed broadband metasurface
	holograms, Nat. Commun. 6~(1) (2015) 8241.
	
	\bibitem{Cheng15b}
	F.~Cheng, J.~Gao, T.~S. Luk, X.~Yang, Structural color printing based on
	plasmonic metasurfaces of perfect light absorption, Sci. Rep. 5~(1) (2015)
	11045.
	
	\bibitem{Sun17}
	S.~Sun, Z.~Zhou, C.~Zhang, Y.~Gao, Z.~Duan, S.~Xiao, Q.~Song, All-dielectric
	full-color printing with {TiO}$_2$ metasurfaces, ACS Nano 11~(5) (2017)
	4445--4452.
	
	\bibitem{Yang20}
	W.~Yang, S.~Xiao, Q.~Song, Y.~Liu, Y.~Wu, S.~Wang, J.~Yu, J.~Han, D.-P. Tsai,
	All-dielectric metasurface for high-performance structural color, Nat.
	Commun. 11~(1) (2020) 1864.
	
	\bibitem{Zhang17}
	X.~Zhang, H.~Liu, L.~Li, Tri-band miniaturized wide-angle and
	polarization-insensitive metasurface for ambient energy harvesting, Appl.
	Phys. Lett. 111~(7) (2017).
	
	\bibitem{Ghaderi18}
	B.~Ghaderi, V.~Nayyeri, M.~Soleimani, O.~M. Ramahi, Pixelated metasurface for
	dual-band and multi-polarization electromagnetic energy harvesting, Sci. Rep.
	8~(1) (2018) 13227.
	
	\bibitem{Li20}
	L.~Li, X.~Zhang, C.~Song, Y.~Huang, Progress, challenges, and perspective on
	metasurfaces for ambient radio frequency energy harvesting, Appl. Phys. Lett.
	116~(6) (2020).
	
	\bibitem{Akselrod15}
	G.~M. Akselrod, J.~Huang, T.~B. Hoang, P.~T. Bowen, L.~Su, D.~R. Smith, M.~H.
	Mikkelsen, Large-area metasurface perfect absorbers from visible to
	near-infrared, Adv. Mater. 27~(48) (2015) 8028--8034.
	
	\bibitem{Raad20}
	S.~H. Raad, Z.~Atlasbaf, C.~J. Zapata-Rodr{\'\i}guez, Broadband absorption
	using all-graphene grating-coupled nanoparticles on a reflector, Sci. Rep.
	10~(1) (2020) 19060.
	
	\bibitem{Karimi22}
	M.~Karimi~Habil, M.~Ghahremani, C.~J. Zapata-Rodr{\'\i}guez, Multi-octave
	metasurface-based refractory superabsorber enhanced by a tapered unit-cell
	structure, Sci. Rep. 12~(1) (2022) 17066.
	
	\bibitem{Jahani16}
	S.~Jahani, Z.~Jacob, All-dielectric metamaterials, Nat. nanotechnol. 11~(1)
	(2016) 23--36.
	
	\bibitem{Kamali18}
	S.~M. Kamali, E.~Arbabi, A.~Arbabi, A.~Faraon, A review of dielectric optical
	metasurfaces for wavefront control, Nanophotonics 7~(6) (2018) 1041--1068.
	
	\bibitem{Koshelev20}
	K.~Koshelev, Y.~Kivshar, Dielectric resonant metaphotonics, ACS Photonics 8~(1)
	(2020) 102--112.
	
	\bibitem{Liu19}
	Z.~Liu, Y.~Xu, Y.~Lin, J.~Xiang, T.~Feng, Q.~Cao, J.~Li, S.~Lan, J.~Liu, High-q
	quasibound states in the continuum for nonlinear metasurfaces, Phys. Rev.
	Lett. 123~(25) (2019) 253901.
	
	\bibitem{Koshelev19}
	K.~Koshelev, Y.~Tang, K.~Li, D.-Y. Choi, G.~Li, Y.~Kivshar, Nonlinear
	metasurfaces governed by bound states in the continuum, ACS Photonics 6~(7)
	(2019) 1639--1644.
	
	\bibitem{Zograf22}
	G.~Zograf, K.~Koshelev, A.~Zalogina, V.~Korolev, R.~Hollinger, D.-Y. Choi,
	M.~Zuerch, C.~Spielmann, B.~Luther-Davies, D.~Kartashov, et~al.,
	High-harmonic generation from resonant dielectric metasurfaces empowered by
	bound states in the continuum, ACS Photonics 9~(2) (2022) 567--574.
	
	\bibitem{Kim24}
	K.-H. Kim, U.-H. An, Near-unity asymmetric transmission of linearly polarized
	light with high {Q}-factor based on the coherent coupling of dual resonances
	in bilayer dielectric chiral metasurfaces, Opt. Laser Technol. 176 (2024)
	110934.
	
	\bibitem{Wang24}
	Z.~Wang, Y.~Zheng, M.~Ouyang, H.~Fan, Q.~Dai, H.~Liu, L.~Wu, High-{Q}
	quasi-bound states in the continuum in c$_2$-symmetric metasurface with
	enhanced second harmonic generation in two-dimensional materials, Opt. Laser
	Technol. 176 (2024) 110868.
	
	\bibitem{Kodigala17}
	A.~Kodigala, T.~Lepetit, Q.~Gu, B.~Bahari, Y.~Fainman, B.~Kant{\'e}, Lasing
	action from photonic bound states in continuum, Nature 541~(7636) (2017)
	196--199.
	
	\bibitem{Spagele21}
	C.~Sp{\"a}gele, M.~Tamagnone, D.~Kazakov, M.~Ossiander, M.~Piccardo,
	F.~Capasso, Multifunctional wide-angle optics and lasing based on supercell
	metasurfaces, Nat. Commun. 12~(1) (2021) 3787.
	
	\bibitem{Tripathi21}
	A.~Tripathi, H.-R. Kim, P.~Tonkaev, S.-J. Lee, S.~V. Makarov, S.~S. Kruk, M.~V.
	Rybin, H.-G. Park, Y.~Kivshar, Lasing action from anapole metasurfaces, Nano
	Lett. 21~(15) (2021) 6563--6568.
	
	\bibitem{Gorkunov20}
	M.~V. Gorkunov, A.~A. Antonov, Y.~S. Kivshar, Metasurfaces with maximum
	chirality empowered by bound states in the continuum, Phys. Rev. Lett.
	125~(9) (2020) 093903.
	
	\bibitem{Chen21}
	X.~Chen, Y.~Zhou, X.~Ma, W.~Fang, W.~Zhang, W.~Gao, Polarization conversion in
	anisotropic dielectric metasurfaces originating from bound states in the
	continuum, Opt. Lett. 46~(17) (2021) 4120--4123.
	
	\bibitem{Pura22}
	J.~L. Pura, R.~Kabonire, D.~R. Abujetas, J.~A. S{\'a}nchez-Gil, Tailoring
	polarization conversion in achiral all-dielectric metasurfaces by using
	quasi-bound states in the continuum, Nanomaterials 12~(13) (2022) 2252.
	
	\bibitem{Tan21}
	T.~C. Tan, Y.~K. Srivastava, R.~T. Ako, W.~Wang, M.~Bhaskaran, S.~Sriram,
	I.~Al-Naib, E.~Plum, R.~Singh, Active control of nanodielectric-induced {THz}
	quasi-{BIC} in flexible metasurfaces: a platform for modulation and sensing,
	Adv. Mater. 33~(27) (2021) 2100836.
	
	\bibitem{Wang21}
	Y.~Wang, Z.~Han, Y.~Du, J.~Qin, Ultrasensitive terahertz sensing with high-{Q}
	toroidal dipole resonance governed by bound states in the continuum in
	all-dielectric metasurface, Nanophotonics 10~(4) (2021) 1295--1307.
	
	\bibitem{Liu23}
	B.~Liu, Y.~Peng, Z.~Jin, X.~Wu, H.~Gu, D.~Wei, Y.~Zhu, S.~Zhuang, Terahertz
	ultrasensitive biosensor based on wide-area and intense light-matter
	interaction supported by {QBIC}, Chem. Eng. J. 462 (2023) 142347.
	
	\bibitem{Hong24}
	W.~Hong, S.~Liu, X.~Sui, X.~Hu, W.~Gu, Ultra-sensitive refractive index sensor
	based on quasi-{BIC} formed in rectangular-split solid-ring metasurface with
	thin film lithium niobate, Opt. Laser Technol. 175 (2024) 110776.
	
	\bibitem{Fleischhauer05}
	M.~Fleischhauer, A.~Imamoglu, J.~P. Marangos, Electromagnetically induced
	transparency: Optics in coherent media, Rev. Mod. Phys. 77~(2) (2005) 633.
	
	\bibitem{Papasimakis08}
	N.~Papasimakis, V.~A. Fedotov, N.~Zheludev, S.~Prosvirnin, Metamaterial analog
	of electromagnetically induced transparency, Phys. Rev. Lett. 101~(25) (2008)
	253903.
	
	\bibitem{Abujetas21}
	D.~R. Abujetas, {\'A}.~Barreda, F.~Moreno, A.~Litman, J.-M. Geffrin, J.~A.
	S{\'a}nchez-Gil, High-q transparency band in all-dielectric metasurfaces
	induced by a quasi bound state in the continuum, Laser Photonics Rev. 15~(1)
	(2021) 2000263.
	
	\bibitem{He21}
	F.~He, J.~Liu, G.~Pan, F.~Shu, X.~Jing, Z.~Hong, Analogue of
	electromagnetically induced transparency in an all-dielectric double-layer
	metasurface based on bound states in the continuum, Nanomaterials 11~(9)
	(2021) 2343.
	
	\bibitem{Zheng24}
	H.~Zheng, Y.~Zheng, M.~Ouyang, H.~Fan, Q.~Dai, H.~Liu, L.~Wu,
	Electromagnetically induced transparency enabled by quasi-bound states in the
	continuum modulated by epsilon-near-zero materials, Opt. Express 32~(5)
	(2024) 7318--7331.
	
	\bibitem{Algorri22}
	J.~Algorri, F.~Dell’Olio, P.~Rold{\'a}n-Varona, L.~Rodr{\'\i}guez-Cobo,
	J.~L{\'o}pez-Higuera, J.~S{\'a}nchez-Pena, V.~Dmitriev, D.~Zografopoulos,
	Analogue of electromagnetically induced transparency in square slotted
	silicon metasurfaces supporting bound states in the continuum, Opt. Express
	30~(3) (2022) 4615--4630.
	
	\bibitem{Huang22}
	L.~Huang, H.~Li, S.~Yu, T.~Zhao, Analogue of electromagnetically induced
	transparency inspired by bound states in the continuum and toroidal dipolar
	response in all-dielectric metasurfaces, Photonics Nanostructures - Fundam.
	Appl. 51 (2022) 101041.
	
	\bibitem{Zhang08c}
	S.~Zhang, D.~A. Genov, Y.~Wang, M.~Liu, X.~Zhang, Plasmon-induced transparency
	in metamaterials, Phys. Rev. Lett. 101~(4) (2008) 047401.
	
	\bibitem{Liu09}
	N.~Liu, L.~Langguth, T.~Weiss, J.~K{\"a}stel, M.~Fleischhauer, T.~Pfau,
	H.~Giessen, Plasmonic analogue of electromagnetically induced transparency at
	the drude damping limit, Nat. Mater. 8~(9) (2009) 758--762.
	
	\bibitem{Han19}
	S.~Han, L.~Cong, Y.~K. Srivastava, B.~Qiang, M.~V. Rybin, A.~Kumar, R.~Jain,
	W.~X. Lim, V.~G. Achanta, S.~S. Prabhu, et~al., All-dielectric active
	terahertz photonics driven by bound states in the continuum, Adv. Mater.
	31~(37) (2019) 1901921.
	
	\bibitem{Han21}
	S.~Han, P.~Pitchappa, W.~Wang, Y.~K. Srivastava, M.~V. Rybin, R.~Singh,
	Extended bound states in the continuum with symmetry-broken terahertz
	dielectric metasurfaces, Adv. Opt. Mater. 9~(7) (2021) 2002001.
	
	\bibitem{Chai21}
	R.~Chai, W.~Liu, Z.~Li, H.~Cheng, J.~Tian, S.~Chen, Multiband quasibound states
	in the continuum engineered by space-group-invariant metasurfaces, Phys. Rev.
	B 104~(7) (2021) 075149.
	
	\bibitem{You23}
	S.~You, M.~Zhou, L.~Xu, D.~Chen, M.~Fan, J.~Huang, W.~Ma, S.~Luo, M.~Rahmani,
	C.~Zhou, et~al., Quasi-bound states in the continuum with a stable resonance
	wavelength in dimer dielectric metasurfaces, Nanophotonics 12~(11) (2023)
	2051--2060.
	
	\bibitem{Ghahremani23}
	M.~Ghahremani, M.~Shahabadi, Accurate characterization of complex bloch modes
	in optical chain waveguides using real-valued computations, Sci. Rep. 13~(1)
	(2023) 22115.
	
	\bibitem{Cai21}
	Y.~Cai, Y.~Huang, K.~Zhu, H.~Wu, Symmetric metasurface with dual band
	polarization-independent high-q resonances governed by symmetry-protected
	bic, Opt. Lett. 46~(16) (2021) 4049--4052.
	
	\bibitem{Fan22}
	H.~Fan, J.~Li, C.~Liu, Y.~Sun, Y.~Wang, X.~Wang, T.~Wu, H.~Ye, Y.~Liu,
	Polarization-independent tetramer metasurface with {multi-Fano} resonances
	based on symmetry-protected bound states in the continuum, Opt. Commun. 525
	(2022) 128864.
	
	\bibitem{Wang21b}
	J.~Wang, J.~K{\"u}hne, T.~Karamanos, C.~Rockstuhl, S.~A. Maier, A.~Tittl,
	All-dielectric crescent metasurface sensor driven by bound states in the
	continuum, Adv. Funct. Mater. 31~(46) (2021) 2104652.
	
	\bibitem{Hsiao22}
	H.-H. Hsiao, Y.-C. Hsu, A.-Y. Liu, J.-C. Hsieh, Y.-H. Lin, Ultrasensitive
	refractive index sensing based on the quasi-bound states in the continuum of
	all-dielectric metasurfaces, Adv. Opt. Mater. 10~(19) (2022) 2200812.
	
	\bibitem{Algorri23}
	J.~Algorri, F.~Dell’Olio, Y.~Ding, F.~Labb{\'e}, V.~Dmitriev, J.~M.
	L{\'o}pez-Higuera, J.~M. S{\'a}nchez-Pena, L.~C. Andreani, M.~Galli, D.~C.
	Zografopoulos, Experimental demonstration of a silicon-slot quasi-bound state
	in the continuum in near-infrared all-dielectric metasurfaces, Opt. Laser
	Technol. 161 (2023) 109199.
	
	\bibitem{Zhou23}
	Y.~Zhou, M.~Luo, X.~Zhao, Y.~Li, Q.~Wang, Z.~Liu, J.~Guo, Z.~Guo, J.~Liu,
	X.~Wu, Asymmetric tetramer metasurface sensor governed by quasi-bound states
	in the continuum, Nanophotonics 12~(7) (2023) 1295--1307.
	
	\bibitem{Li23}
	Z.~Li, M.~Panmai, L.~Zhou, S.~Li, S.~Liu, J.~Zeng, S.~Lan, Optical sensing and
	switching in the visible light spectrum based on the bound states in the
	continuum formed in {GaP} metasurfaces, Appl. Surf. Sci. 620 (2023) 156779.
	
	\bibitem{Wang23}
	K.~Wang, H.~Liu, Z.~Li, M.~Liu, Y.~Zhang, H.~Zhang, All-dielectric
	metasurface-based multimode sensing with symmetry-protected and accidental
	bound states in the continuum, Results Phys. 46 (2023) 106276.
	
	\bibitem{Sherry05}
	L.~J. Sherry, S.-H. Chang, G.~C. Schatz, R.~P. Van~Duyne, B.~J. Wiley, Y.~Xia,
	Localized surface plasmon resonance spectroscopy of single silver nanocubes,
	Nano Lett. 5~(10) (2005) 2034--2038.
	
	\bibitem{Tian16}
	J.~Tian, Q.~Li, Y.~Yang, M.~Qiu, Tailoring unidirectional angular radiation
	through multipolar interference in a single-element subwavelength
	all-dielectric stair-like nanoantenna, Nanoscale 8~(7) (2016) 4047--4053.
	
	\bibitem{Terekhov17}
	P.~D. Terekhov, K.~V. Baryshnikova, Y.~A. Artemyev, A.~Karabchevsky, A.~S.
	Shalin, A.~B. Evlyukhin, Multipolar response of nonspherical silicon
	nanoparticles in the visible and near-infrared spectral ranges, Phys. Rev. B
	96~(3) (2017) 035443.
	
	\bibitem{Zenin17}
	V.~A. Zenin, A.~B. Evlyukhin, S.~M. Novikov, Y.~Yang, R.~Malureanu, A.~V.
	Lavrinenko, B.~N. Chichkov, S.~I. Bozhevolnyi, Direct amplitude-phase
	near-field observation of higher-order anapole states, Nano Lett. 17~(11)
	(2017) 7152--7159.
	
	\bibitem{Babicheva21}
	V.~E. Babicheva, A.~B. Evlyukhin, Multipole lattice effects in high refractive
	index metasurfaces, J. Appl. Phys. 129~(4) (2021).
	
\end{thebibliography}

\end{document}